# Examining Cashless Payment Services in a "Post" Pandemic Environment

Janiya R. Peters
University of California, Berkeley,
janiya_peters@berkeley.edu

Orissa Rose
University of California, Berkeley,
orissarose@berkeley.edu

The global pandemic COVID-19 posed numerous challenges for U.S. restaurants and food services. Many businesses adopted contactless ordering and cashless payment policies to comply with emergency health mandates. Even with national and public health emergency mandates set to expire in May 2023 [22], cashless payment services continue to thrive through online ordering platforms such as DoorDash and Uber Eats and social payment platforms such as Snackpass. At present, designers and policymakers must address the socioeconomic politics of cashless payment services and service accessibility for marginalized groups.

## 1 INTRODUCTION

In October-December 2021, we conducted a document analysis of 56 store payment policies in Berkeley, CA. During this time, the COVID-19 pandemic forced government restrictions on group gatherings and in-store patronage. We identified food services which employed cashless payment options, including mobile ordering applications, self-service kiosks, and contactless pickup trays. Our photo documents revealed obstruction of store entry, reduced access to cash registers or queues, public safety concerns and currency limitations.

This comes at a time when Berkeley reports 1,000+ unhoused residents, many of whom do not have digital payment options. We identified Berkeley Municipal Code 9.50.030, stating: "Except as set forth in 9.50.040, every Covered Business within the City must accept payment in Cash, if offered, for any transaction involving the purchase of any tangible good and/or service" [7] [Appendix B]. Failure to comply means low-income, unbanked, and unhoused persons face reduced access to food and local services. In May 2022, we addressed these concerns to the Berkeley City Attorney's Office in a joint letter with the American Civil Liberties Union Foundation (ACLU) of Northern California, Housing and Economics Rights Advocates (HERA), and East Bay Community Law Center. In October 2022, the City responded with an invitation to pursue legal recourse ourselves, but offered no commitment to City-led policy enforcement.

## 2 METHODS

We photographed and documented store payment policies for 56 businesses and restaurants within Berkeley, CA. At this time, UC Berkeley students and faculty were returning to campus and restaurants were consistently open to meet demand; this allowed us to see a real-time response to pandemic and economic conditions within a concentrated area. Businesses were selected within a one-mile radius of UC Berkeley's Sproul Hall, comprised of restaurants, food services, and a retail pharmacy.

We categorized two types of documents: **explicit payment policies** and **implicit payment policies**. Explicit payment policies include print signage: window posters, stickers, adverts, plaques, or any paper documents that display textual information about payment protocols. Print signage is a direct, straight-forward means of communication between business and customer. Implicit payment policies appeared through infrastructure: any physical artifact, object, or structure indicating method of payment. Infrastructure includes self-service kiosks, card readers, cash registers, coin machines, pick-up tray stations and physical barricades. Each infrastructural object carries its own affordances based on its physical properties which signal to the user what actions are possible (and, by design, which are not) [10, 18]. For example, a card reader affords a particular swiping motion of a flat, thin plastic through the slit. Affordances may be shaped by environment the object is placed in. For example, a chair affords the act of sitting; but multiple chairs placed in front on a queue might indicate a stoppage of motion where the patron is unable to proceed.

Finally, we coded the documents for payment status (Table 1) and recurring themes (Table 2). Payment status was singularly determined based on photographed documents, including print signage and infrastructure which is openly visible to the customer. A single store could be labeled with multiple store policy codes. For example, a store may be labeled *mobile ordering* and *contactless* based on multiple photo documents.

Table 1: Payment Status

| PAYMENT STATUS | DESCRIPTION |
|---|---|
| CASH | Physical dollars, coins, and bank notes accepted or preferred, as indicated by signage and/or infrastructure. **Cash-friendly.** |
| DIGITAL PAYMENT | Credit card, debit card, mobile wallets, and other electronic payments accepted or preferred, as indicated by signage and/or infrastructure. **Cashless or cash heavily restricted.** |
| MIXED PAYMENT | Both cash and digital payment are accepted, as indicated by signage and/or infrastructure. |
| NOT CLEAR | Not clear or cannot be determined based on documents whether cash or digital payment is preferred. |

Table 2: Store Policy Codes

| CODES | DESCRIPTION | EXEMPLAR |
|---|---|---|
| COIN SHORTAGE | Lack of physical dollars, coins, or notes discourages cash or restricts the type of cash accepted. | "Due to the national shortage of coins in circulation…" |
| CONTACTLESS | Reduced human contact in the completion of a transaction. | "Contactless pickup," "Contact-free ordering." |
| GRATUITY | Changes to voluntary and involuntary tipping, (including both cash and digital tipping). Tipping is an additional cost to the consumer to their actual purchase. | "We need the dough!" posted on a plastic cup at a pizzeria. |
| ENTRY BLOCKED | Physical obstruction of store entry by object or process. | Blocking the door with a kiosk. |

| CODES | DESCRIPTION | EXEMPLAR |
|---|---|---|
| HEALTH | Norms adopted to protect the physical well-being of customers, staff, and the public. | "Due to the increased cases of COVID…" |
| KIOSK | Presence of a digital self-ordering station where customers can make a purchase. | "Contactless ordering! Self-service kiosk order." |
| MOBILE ORDERING | The facilitation of a smartphone, mobile app, phone camera, or additional device to transact. | Snackpass, DoorDash. |
| REGISTER | Presence of a cash drawer to make transactions. | Item from which money is removed or deposited. Operated by cashier. |

## 3 DOCUMENT ANALYSIS

Based on signage and infrastructure present, businesses in our study highly preferred digital payment. Seven stores welcomed mixed payment and one store indicated they only accept cash (Table 3). When coding store policies, we allowed for a single store to have up to four codes (Table 4). For example, a convenience store might show instances of a *register* and *coin shortage* according to a single or multiple photo documents. The most recurring theme in our documents was *mobile ordering*; this was practical given that our sampled stores were primarily restaurants and food services. Mobile ordering apps like Snackpass and DoorDash appeared frequently in these documents.

Table 3: Payment Status Count

|  | CASH | DIGITAL PAYMENT | MIXED PAYMENT | NOT CLEAR |
|---|---|---|---|---|
| COUNT | 1 | 43 | 7 | 5 |
| PERCENTAGE | 1.8% | 76.8% | 12.5% | 8.9% |

Table 4: Store Policy Code Count

|  | COIN SHORTAGE | CONTACT-LESS | ENTRY BLOCKED | GRATUITY | HEALTH | KIOSK | MOBILE ORDERING | REGISTER |
|---|---|---|---|---|---|---|---|---|
| COUNT | 3 | 9 | 5 | 7 | 2 | 17 | 38 | 3 |

### 3.1 Signage Indicates Monetary and Health Concerns

In three cases, signage or posted policy referred to a *coin shortage*. These notices proclaimed: "Out of change! Please pay exact change or card" and "We are temporarily unable to accept cash for your order unless you have the exact change. We can accept credit, debit, and gift cards. We apologize for the inconvenience." One pharmacy stated: "due to the national shortage of coins in circulation to pay for goods and services, we encourage the use of electronic payments or exact change when possible." This signage communicated a lack of dollars and/or coins on hand, potentially as an economic side effect of COVID-19. Due to closure of businesses and services during quarantine, the circulation of physical coins slowed considerably. When establishments reopened, stores experienced a higher demand for coins than what they had in supply [19]. This explains the preference for "exact change when possible." The scale of the coin shortage varies by establishment. While two stores posted that only their location was out of change, the pharmacy invoked the term "national shortage," indicating the problem extended past their store location.

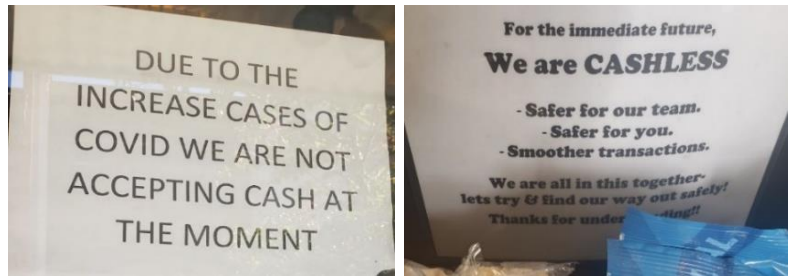

Figure 1: Signage provides health and safety concerns as reason for adopting cashless policies.

Two stores that did not accept cash cited COVID-19 and communal safety for changes to payment policy. We labeled these documents as *health* concerns. The first establishment rationalized the change as a response to increased COVID-19 cases; the second establishment posted they were "cashless" because it was "safer for our team," "safer for you," and yielded "smoother transactions" (Figure 1).

The first policy fails to state where COVID-19 cases increased, whether locally, statewide, nationally, etc. The second policy does not explicitly reference COVID-19, but it is strongly implied in the context of several pandemic responses and emergency mandates. In place of cash, the establishments offer customers to pay by card, Snackpass kiosk or Snackpass mobile app (use of Snackpass services requires a debit or credit card). However, it is unclear how frequently staff sanitize kiosks or card readers.

The case for cashless policies amid COVID-19 sparked major debate. On March 2, 2020, UK news outlet The Telegraph reported that World Health Organization (WHO) spokeswoman Fandela Chaib warned people against the exchange of banknotes, recommending contactless payments as a substitute [9]. Several outlets have since clarified Chaib's claim, stating:

> "WHO did NOT say banknotes would transmit COVID-19, nor have we issued any warnings or statements about this," Chaib said in an email. "We were asked if we thought banknotes could transmit COVID-19 and we said you should wash your hands after handling money, especially if handling or eating food." Doing so is "good hygiene practice," she added. [11]

Despite the clarification, digital payment became a preferred, "safer" payment method amid uncertain guidelines and contrary narratives.

## 3.2 How Infrastructure Limits Payment Options

We noted several cases of infrastructure physically blocking areas of entry by way of kiosks, tables, and chairs. These documents were coded as *kiosk* wherever the device was present and/or *entry blocked* wherever the kiosk (or another mechanism) was used to deter entry to an area of the store. One store posted a sign saying, "No Cash" and blocked the door with a table and kiosk. A second store blocked access to the cash register with a row of chairs; tacked onto these chairs was a sign with QR code instructions to order online.

The use of infrastructure to block entry physically changes and limits customers' payment options. Through spatial manipulation, the customer is defaulted to interact with the kiosk. These decisions adversely impact individuals lacking cards and/or smartphones. Ordering kiosks require a phone number or email to transact; ordering online requires a phone or computer; and QR codes require a camera-enabled phone to scan and access menu and further services. These items are not frequently accessible for unbanked, elderly, low-income, and homeless populations.

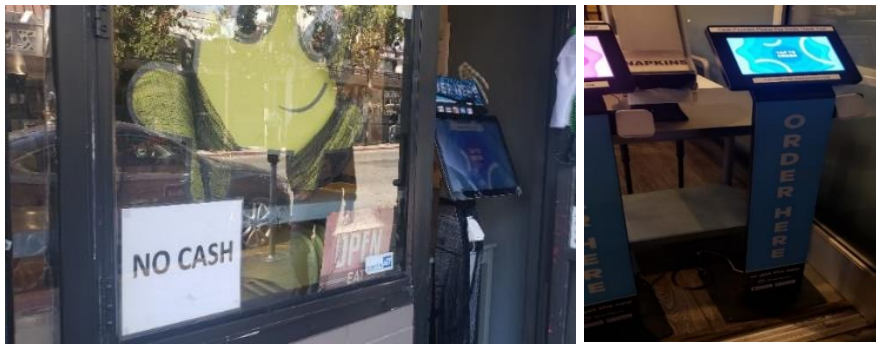

Figure 2: Tables and Snackpass kiosks block entryway into restaurants.

The kiosk is an essential component in cashless payment services. Customers can view item options, add/remove items, input mobile contact numbers, and insert cards into attached card readers. However, the rise of kiosks is accompanied by lawsuits arguing that they are not accessible to wheelchair users and individuals with vision and hearing impairments [15]. The following characteristics of an ADA compliant kiosk were noticeably absent from the kiosks we photographed: unobstructed wheelchair access both forward, parallel, front and back; is not too high up or too far back on the counter; is not too small; offers verbal assistance as needed; no protruding objects or wall mounts that would impede the path of the user or be unexpected by individuals who are vision-impaired or using a cane; has clear sign indicating that it's an ADA compliant kiosk; interface design places elements in lower positions of the screen [12].

In Figure 2, the kiosk accommodates standing customers. There are no visible height adjustments. Individuals who use wheelchairs might have difficulty seeing the screen or getting close enough due to the kiosk's position in the door frame. There is no sign indicating that the kiosk is ADA compliant. In the right image, there is a label on the kiosk indicating "Cash Payment Please Pay Inside Thank You!" However, due to the blocked entrance, it is unclear how a customer might enter the store to pay. This setup creates accessibility issues for the public, especially elderly persons and individuals with disabilities. Without an employee present to provide customer support, options for clarification are limited.

### 3.3 The Branding of "Contactless"

Nine stores indicated a *contactless* ordering system, which reduces human contact in the transaction process. "Contact free," "contactless payments," "contactless pickup," and "touch free" were recurring words in our documents. *Contactless* ordering systems were frequently accompanied by *mobile ordering,* which appeared 38 times in our sample, and QR ordering instructions [Appendix A]. We reasoned that because contactless ordering requires customers to order and pay in advance, mobile ordering would naturally accompany this mechanism through apps like Uber Eats, DoorDash, and Snackpass. Social distancing mandates, including the suggested "6 feet apart" distance rule, prompted many food establishments to adopt contactless ordering in order to continue service. While framed as an act of caution in accordance with health and safety guidelines, it is also a convenient mechanism for customers with digital payment options and businesses interested in streamlining operations. The ability to order and pay online, and pickup when ready, reduces human contact, wait times, number of patrons in-store, as well as the number of staff needed to interact with customers and handle money.

### 3.4 Gratuity

Reduced human contact during COVID-19 yields adverse consequences for gratuity recipients. As cashless systems become normalized, employees across service industries report a decrease in cash tips [8]. In our documents, we observed various *gratuity* mechanisms, including physical jars of money and kiosk requests. In instances when the kiosk prompted customers to tip, the store clarified that "100% of your tip goes to [establishment] staff" [Appendix A]. This statement is necessary to clarify who receives gratuity and/or how the tip will be distributed, especially given reduced staff presence. Some stores retained physical tip jars on the checkout counter. The presence of these jars in stores where cash was *not* accepted was interesting; despite "cashless" signage, these sites still had hard currency visible in their tip jars. The contradiction between payment policies and gratuity expectations may signal a disconnect between management directives and health protocols, and the economic desires of service workers.

## 4 RELATED WORK

The rapid growth of ubiquitous digital technologies, particularly the smartphone, has granted "over 1 billion people to transact via different mobile payment apps worldwide," and quick, convenient access to banking institutions [3]. For formal institutions, digital payments offer regulability; make taxes and formal labor arrangements traceable; and reduce underground economic activities [13]. Discussions of a future "cashless society" in which all monetary exchanges are conducted "electronically" without checks or paper money can be traced as far back as 1959 [6]. However, immigrant and elderly populations, and individuals with physical and cognitive disabilities rely on cash for daily exchange [4]. Even as digital food ordering platforms offer streamlined payment options, they risk excluding more vulnerable populations from their service model. Social payment platforms (e.g., Venmo, Snackpass) steepen this divide and introduce new privacy risks by incorporating social media features such as posts, feeds, and user profiles [1, 2]. These features find mass appeal in college campus communities [14], privileging a younger, technologically equipped consumer base.

## 5 RECOMMENDATIONS

Cashless payment services function as a *technical arrangement* which initiates a *form of order:* an order that streamlines services for individuals with sufficient banking and technological resources [21]. Recognizing the health and economic *triggers* for payment reconfigurations in Berkeley in 2021, we propose a model of commerce that returns agency to customers and employees *displaced* by cashless models [17]. We adopt Chelsea Barabas' "Ethical AI" approach in order to: attribute agency and responsibility where it is due; reframe utopian narratives about cashless societies that exclude marginalized groups; and define fair practices between community stakeholders [5]. We encourage practitioners, policymakers, and businesses adopting cashless payment services to:

- Inform customers of alternate payment options through signage and maintain entry to cash infrastructure. This means keeping queues open and accessible, and maintaining sufficient legal tender to complete cash transactions.
- Maintain ADA compliant kiosks with height adjustments, voice and visual assistance. Kiosks *should not* block doors or entryways.
- Embed disclosures into digital ordering services indicating the recipient of customers' tips and additional fees.
- Annually audit store payment infrastructures for compliance with local commercial ordinances.

We remain vigilant of the continued risks COVID-19 poses to our global community. We insist upon sociotechnical systems that *both maximize public safety* and *remain accessible to all community members*. Future work may audit cashless payment services across college communities and evaluate them based on principles of fairness and privacy [16, 20].

**ACKNOWLEDGMENTS**

We thank UC Berkeley faculty advisor Deirdre K. Mulligan, and ACLU of Northern California members Nicole A. Ozer and Jacob Snow for their continued feedback, guidance, and advocacy.

**APPENDIX A: Additional Payment Policy Documentation**

**A1. The Branding of "Contactless"**

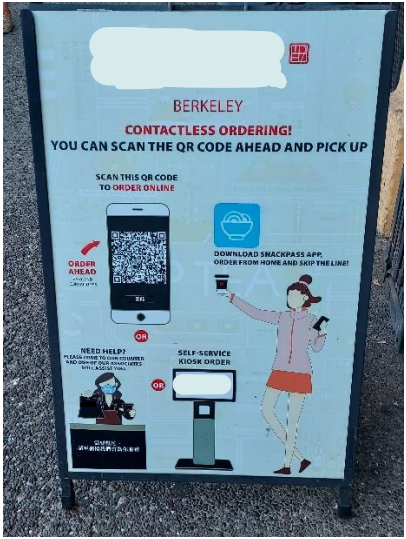

**A2. Gratuity**

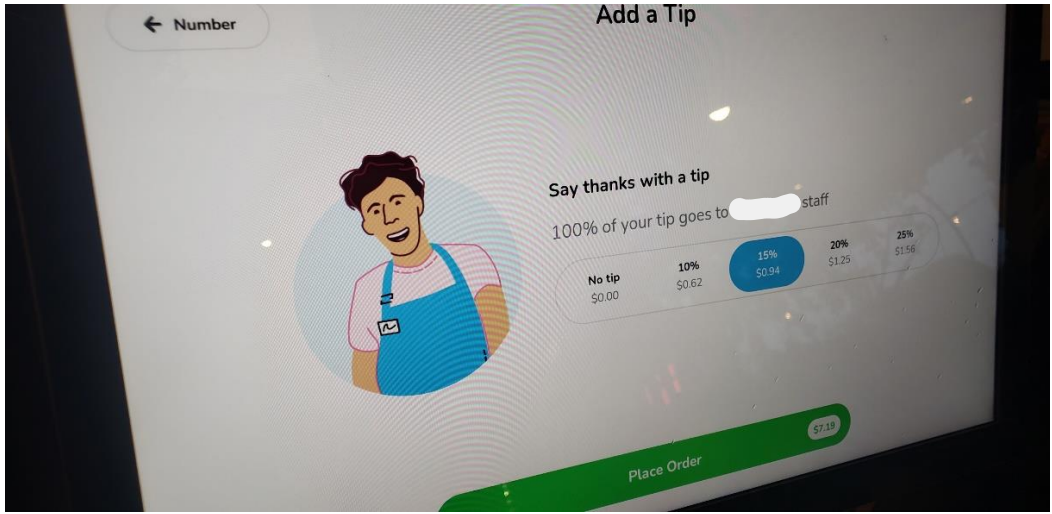

**APPENDIX B: City of Berkeley Municipal Code 9.50**

# Chapter 9.50
# LEGAL RIGHTS FOR LEGAL TENDER

Sections:

| | |
|---|---|
| 9.50.010 | Findings and purpose. |
| 9.50.020 | Definitions. |
| 9.50.030 | Covered Businesses Required to Accept Cash. |
| 9.50.040 | Exceptions. |
| 9.50.050 | Enforcement. |
| 9.50.060 | Severability. |

## 9.50.010   Findings and purpose.

The Council finds and declares as follows:

A.   The City of Berkeley is committed to providing its community with transactional access to the goods and services provided by Berkeley's businesses. For many City residents, such as those unable to obtain bank accounts, the ability to engage in consumer transactions, including goods and services vital to health and safety, depends on the ability to pay with legal cash tender established by the federal government of United States.

B.   Cashless business models present significant detrimental impacts to vulnerable groups, especially low-income people, as they require financial institution-sponsored payment in credit or debit cards, or other non-cash forms of payment.

C.   Cash payment, in the form of the United States Dollar, has been the official legal tender since 1792 and shall be recognized by businesses alongside other forms of legal tender.

D.   It is the intent of the Council to ensure Berkeley's economy is inclusionary and accessible to everyone, including those who lack access to non-cash forms of payment. (Ord. 7681-NS § 1 (part), 2019)

## 9.50.020   Definitions.

A.   Covered Business shall mean any Drugstore, Food Products Store, or Retail Products Store operating at a fixed, permanent, physical premises. Covered businesses do not include any transactions occurring in an Itinerant Restaurant as defined in BMC 12.04.010.

B.   "Cash" means United States currency, in the form of both paper Federal Reserve Notes and metal coins.

C.   "Drugstore" shall have the same meaning as defined in BMC 23F.04.010.

D.   "Food Products Store" shall have the same meaning as defined in BMC 23F.04.010.

E.    "Retail Products Store" shall have the same meaning as defined in BMC 23F.04.010. (Ord. 7681-NS § 1 (part), 2019)

### 9.50.030     Covered Businesses Required to Accept Cash.

A.    Except as set forth in 9.50.040, every Covered Business within the City must accept payment in Cash, if offered, for any transaction involving the purchase of any tangible good and/or service.

B.    Except as set forward in 9.50.040, a Covered Business may not charge a fee or place any other condition on its acceptance of Cash as required by subsection A. (Ord. 7681-NS § 1 (part), 2019)

### 9.50.040     Exceptions.

The provisions set forward in this Act shall not apply in cases of:

A.    *Suspected counterfeit currency.* A Covered Business may refuse to accept Cash that the business reasonably suspects to be counterfeit.

B.    *Large denominations.* A Covered Business may refuse to accept Cash in any denomination larger than a twenty dollar note, but shall otherwise accept any combination of Federal Reserve Notes and metal coins in connection with any transaction.

C.    *Single transactions above $500.* Where a single transaction involves the purchase of one or more goods and/or services, the total price of which (including tax) exceeds $500, a Covered Business must accept Cash that is offered as payment for any amount up to and including $500, but may refuse to accept Cash that is offered as payment for the remainder of the amount due.

D.    *Reservations made without cash.* Where a Covered Business requires the purchaser make an appointment or reservation using a noncash form of payment (such as a credit or debit card), the business may require that the transaction in question be paid for using the noncash payment already on file. (Ord. 7681-NS § 1 (part), 2019)

### 9.50.050     Enforcement.

A.    The obligation to ensure that a Covered Business complies with this Chapter 9.50 shall fall only on the business or, in the case that the owners of the business are responsible for a policy or practice causing a violation of this Chapter, on the owner or owners of the business. No employee or independent contractor working at a Covered Business shall be held liable for any violation of this Chapter.

B.    Each transaction or attempted transaction in which a Covered Business fails to accept Cash shall constitute a separate violation of this Chapter.

C.  Any aggrieved person who believes the provisions of this Chapter have been violated shall have the right to file an action for injunctive relief and/or damages. In any action to enforce the provisions of the chapter, the prevailing party shall be entitled to recover reasonable attorneys' fees and costs.

D.  The City may issue an Administrative Citation pursuant to Chapter 1.28 of the Berkeley Municipal Code for any violation of this Chapter. The amount of this fine shall be determined as specified below:

   a.  For a first violation, an infraction punishable by a fine not exceeding $100 and not less than $50.

   b.  For a second violation within a twelve month period, an infraction punishable by a fine not exceeding $200 and not less than $100.

   c.  For a third violation within a twelve month period, an infraction punishable by a fine not exceeding $1,000 and not less than $500. (Ord. 7681-NS § 1 (part), 2019)

## 9.50.060    Severability.

If any word, phrase, sentence, part, section, subsection, or other portion of this Chapter, or any application thereof to any person or circumstance is declared void, unconstitutional, or invalid for any reason, then such word, phrase, sentence, part, section, subsection, or other portion, or the prescribed application thereof, shall be severable, and the remaining provisions of this Chapter, and all applications thereof, not having been declared void, unconstitutional or invalid, shall remain in full force and effect. The City Council hereby declares that it would have passed this title, and each section, subsection, sentence, clause and phrase of this Chapter, irrespective of the fact that any one or more sections, subsections, sentences, clauses or phrases is declared invalid or unconstitutional. (Ord. 7681-NS § 1 (part), 2019)

---